\documentclass[twocolumn,aps,prl,longbibliography]{revtex4-1}
\usepackage{amsmath,latexsym}
\usepackage{amssymb}
\usepackage{xcolor}
\usepackage[%
  colorlinks=true,
  urlcolor=blue,
  linkcolor=blue,
  citecolor=blue
]{hyperref}
\usepackage{graphicx}
\usepackage[version=3]{mhchem} 

\makeatletter
\let\cat@comma@active\@empty
\makeatother

\begin{document}

\title{Magnetic-Field-Induced Transition in a Quantum Dot Coupled to a Superconductor}
\author{A. Garc\'ia Corral$^{1}$, D. M. T. van Zanten$^{1,*}$, K. J. Franke$^{2}$, H. Courtois$^{1}$, S. Florens$^{1}$, and C. B. Winkelmann$^{1}$}
\affiliation{$^{1}$\mbox{Univ.} Grenoble Alpes, CNRS, Grenoble INP, Institut N\'eel, 38000 Grenoble, France}
\affiliation{$^{2}$Fachbereich Physik, Freie Universit\"at Berlin, 14195 Berlin, Germany}
\affiliation{$^{*}$present address: Center for Quantum Devices, Niels Bohr Institute,
University of Copenhagen, and Microsoft Quantum Lab Copenhagen, Denmark}
\begin{abstract}
The magnetic moment of a quantum dot can be screened by its coupling to a
superconducting reservoir, depending on the hierarchy of the superconducting gap
and the relevant Kondo scale. This screening-unscreening transition can be 
driven by electrostatic gating, tunnel coupling, and, as we demonstrate here,
magnetic field. We perform high-resolution spectroscopy of subgap
excitations near the screening-unscreening transition of asymmetric superconductor -
quantum dot - superconductor (S--QD--S) junctions formed by the electromigration
technique.
Our measurements reveal a re-entrant phase boundary determined by the competition 
between Zeeman energy and gap reduction with magnetic field. We further track the 
evolution of the phase transition with increasing temperature, which is also 
evidenced by thermal replicas of subgap states.
\end{abstract}

\keywords{quantum dot, superconductivity, Anderson impurity model, Kondo effect}

\maketitle

The junction between a superconductor and a quantum dot displays discrete subgap
energy levels called Andreev bound states (ABS) or, more specifically,
Yu-Shiba-Rusinov (YSR) states when the highest occupied level hosts a
single spin \cite{Balatsky2006,Heinrich2018}. When the antiferromagnetic exchange interaction between this 
spin and the leads prevails over the superconducting gap $\Delta$, the
localized spin is effectively Kondo screened, and a quasiparticle is bound at 
the interface. In contrast, if the exchange coupling is weaker, the superconducting
condensate is marginally perturbed and the spin remains unscreened. The change between these two distinct ground states occurs via a sharp level
crossing, which constitutes a simple realization of a quantum phase transition.
In recent years, detailed investigations of this screening-unscreening
transition have been performed, using as a control knob the variation of the level depth
\cite{Deacon2010,Pillet2010,Lee2014,Jellinggaard2016,Assouline2017} or the tunnel coupling,
which effectively modifies the exchange coupling strength
\cite{Franke2011,Island2017,Farinacci2018,Malavolti2018}. The variation of an external
magnetic field may provide an additional parameter with twofold consequences:
increasing the magnetic field suppresses superconducting correlations (hence
favoring a screened state), while the Zeeman effect enhances polarization towards 
one of the magnetic YSR states \cite{Lee2014, Jellinggaard2016, Li2017, Cornils2017}. 
Thus, the two effects associated to a magnetic field lead to a shifting of level 
crossing in opposite directions.

Experimentally, it is challenging to explore the effect of the magnetic field 
on the screening-unscreening transition, precisely for the reason that
superconductivity is usually quenched before the quantum phase transition can be
accessed. Hence, the quantum dot systems needs to be tuned close to the
critical point using another control parameter, here the back gate
voltage allowed in our transistor geometry. Furthermore, the detection of 
tiny level shifts between subgap states requires exquisite energy 
resolution in the $\mu$V range, which can be achieved only with superconducting
leads.

In this Letter, we report on the observation of the magnetic field-controlled
screened-unscreened ground state transition of a quantum dot strongly coupled 
to one superconducting lead. An asymmetrically coupled superconductor--quantum 
dot--superconductor (S-QD-S) device combines the gate tunability of single 
electron transistors with the unprecedented spectroscopic resolution of the 
subgap states.  Monitoring the
dispersion of the subgap states allows tracking the transition between the
screened and unscreened spin ground states of the quantum dot as a function of
the bare level of the dot, temperature, and, most importantly, magnetic field. 
We use the Anderson impurity as the main framework for the modelization of our data. 
A general phase diagram is drawn, which demonstrates a striking re-entrant
behavior of the phase boundary, due to the previously mentioned competition 
between Zeeman splitting and superconducting gap closing. In addition,
thermal replicas of YSR states are found to emerge at finite temperature,
providing an alternative yet consistent picture of the subgap spectrum.

\begin{figure}[ht]
	\includegraphics[width=1.0\columnwidth]{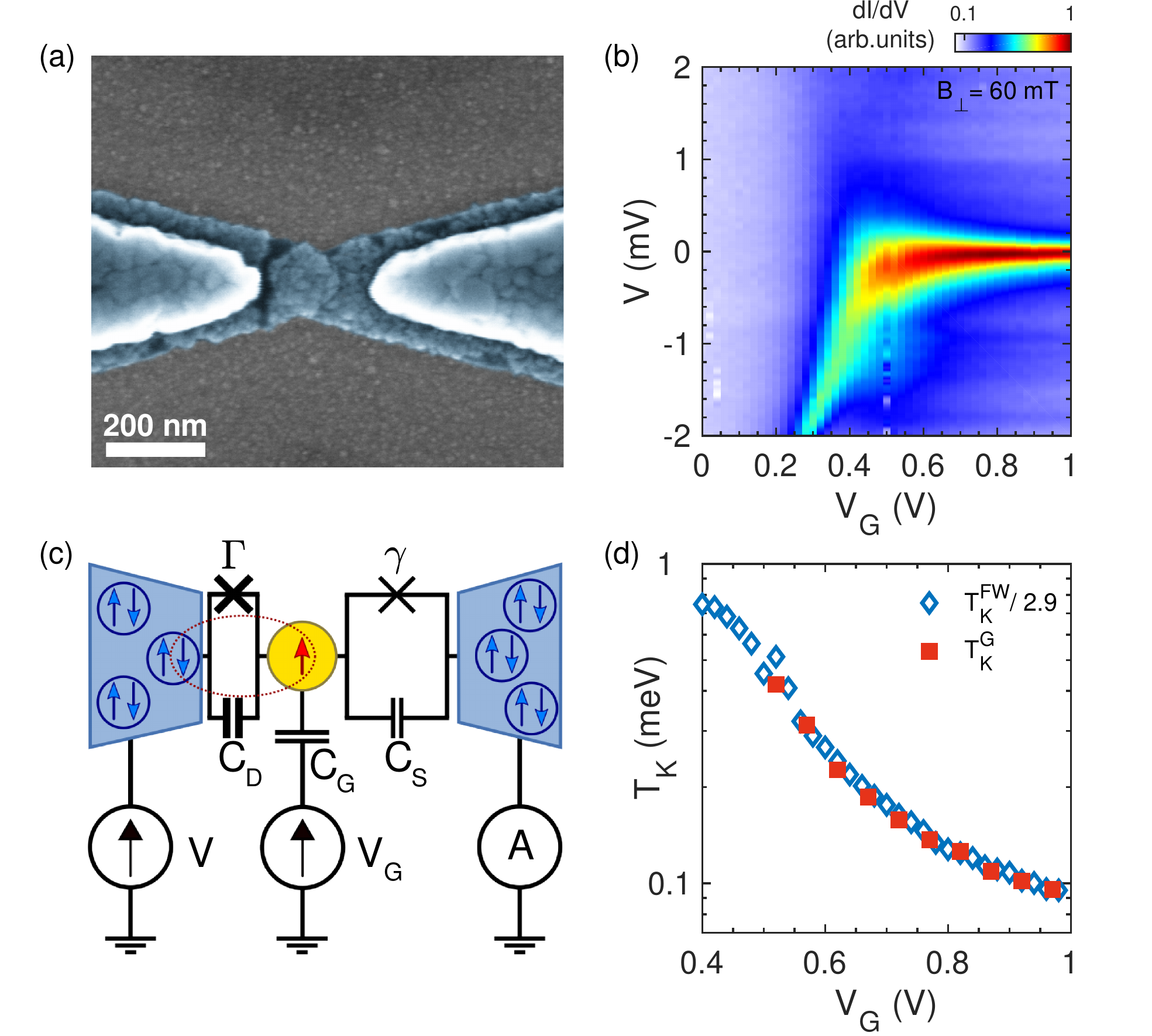}
	\vspace{-0.5cm}
	\caption{(a) Scanning electron micrograph of a bare Al constriction after
electromigration. 
(b) Normal state $dI/dV$ differential conductance map of device A at a base
temperature of $T=100$ mK, for a magnetic field $B=60$ mT suppressing
superconductivity. 
(c) Schematics of the S--QD--S device, introducing the three capacitances and two 
tunnel couplings at play.
(d) Experimental gate dependence of the Kondo energy scale $ T_K$, determined from the 
temperature dependence of the linear conductance (red squares), as well as by a
rescaling with a dividing factor 2.9 of the FWHM of the low-bias conductance peak 
(blue open diamonds). 
}
	\label{normal}
\end{figure} 

The device fabrication process relies on controlled electromigration of an
on-chip all-metallic aluminum device presenting a constriction \cite{Park99}.
This technique produces nanometer sized gaps and was successfully applied for
connecting single molecules \cite{Park00, Park02,Liang02,Vincent2012}.
Electrostatic gate
control is provided through a local metallic back gate isolated by a 18 nm
thick Zr0$_2$ dielectric layer. Using aluminum as the constriction material,
gated S--QD--S devices can thereby be
formed~\cite{Winkelmann09,vanZanten2015,vanZanten2016}. Our quantum dots are colloidal gold nanoparticles of 5
nm diameter. Electromigration is performed at 4.2 K in cryogenic vacuum in a
dilution cryostat. A scanning electron micrograph of an Al constriction after
electromigration (without nanoparticles, for better visibility) is shown in 
Fig.~\ref{normal}a.  Samples showing stable gate-dependent conductance features are
further investigated at temperatures down to $T=80$ mK. The differential
conductance $G(V,V_G)=dI/dV$ is
measured using the lock-in technique, as a function of bias $V$ and gate voltage $V_G$. We show here data mostly
from one sample, labelled A. Data from a second and similar sample (B) can be found 
in the Supplemental Material file. 

The normal state differential conductance map (obtained at a magnetic field of
60 mT) is shown in Fig.~\ref{normal}b, around the only experimentally accessible 
degeneracy point at $V_G^0\approx 0.40$ V.
To the left, the linear conductance is suppressed, owing to a Coulomb-blockaded
state with an even electron occupation number $N$. As the gate voltage $V_G$ is increased, 
a zero-bias resonance indicates the onset of Kondo correlations associated to the
spin-1/2 degeneracy of the oddly occupied $N+1$ electron state. The electrical model of the quantum dot junction is displayed in \mbox{Fig.} \ref{normal}c. The tunnel
couplings to both leads are strongly asymmetric ($\Gamma\gg \gamma$), as
evidenced by a non-unitary linear conductance limit, $G(T\to 0)/G_0=  4\Gamma
\gamma/(\Gamma + \gamma)^2 \approx0.013$, with $G_0=2 e^2/h$ (see Supplemental Material file). 
This implies notably that the Kondo resonance builds between the QD and the drain
electrode at experimentally accessible temperatures, the source contact acting 
as a tunnel probe, as it is usually the case for the tip in an STM experiment.
The following values of hybridization $\Gamma=1.4$ meV and Coulomb repulsion $U=12.7$ meV
in the quantum dot are found from fits of the gate-dependent zero-bias conductance 
to Numerical Renormalization Group (NRG) calculations, see Supplemental Material. 
The ratio $U/\Gamma\simeq 9$ shows that the quantum dot is in the strongly correlated 
regime, with some deviations from Kondo scaling.

The full-width at half-maximum (FWHM) of the conduction resonance at the Fermi
level, as shown in Fig.~\ref{normal}b, is often taken as an approximate measure
of the Kondo temperature $T_K$, that we note $T_K^{FW}$ (here and later 
we set $k_B=1$, identifying temperature and energy scales). 
Another direct and precise determination of $T_K$ is achieved considering the measured 
temperature dependence of the linear conductance $G(T)$. The latter can 
be fitted by NRG calculations, or for a lower computational cost,
by an empirical expression \cite{Goldhaber1998,Dutta2019}, leading to the gate-dependent
Kondo temperature denoted $T_K^G$
shown in \mbox{Fig. \ref{normal}d} (see also Supplemental Material). We find that these estimates agree closely within a scaling factor, such that $T_K^G = T_K^{FW}/2.9$. 
Therefore, it is seen that the quantum dot
junction behaves like a single spin-1/2 Kondo impurity, with a gate-tunable
$T_K$ that can be brought to the same order of magnitude as the superconducting 
order parameter of the leads, leading to a standard gate-control of the
screening transition \cite{Bauer2007,Maurand2012}.

\begin{figure}[ht]
	\hspace{1.7cm}
	\includegraphics[width=1.0\columnwidth]{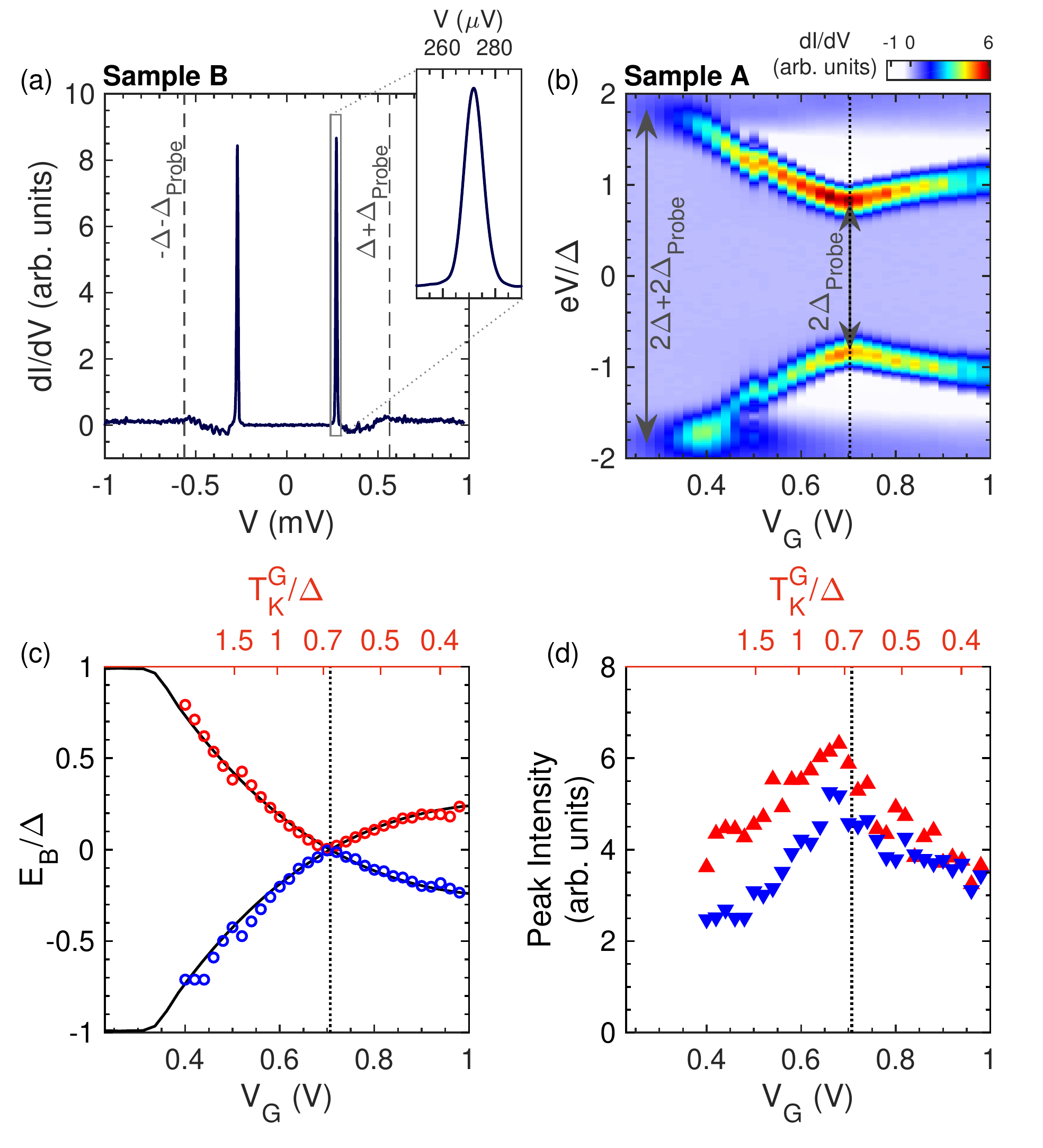}
		\vspace{-0.5cm}
\caption{ (a) Differential conductance of device B measured with a lock-in AC
oscillation of 2 $\mu$V in the superconducting 
state at fixed $V_G=0.5$ V, displaying very sharply resolved
YSR resonances, with a FWHM = 9 $\mu$V (see inset). (b) Gate-dependent
differential conductance of device A in the superconducting state at a base
temperature of $T=80$ mK, 
revealing the dispersion of the YSR states. The minimal spacing of the two resonances,
given by a voltage span $2\Delta_{\rm probe}/e$ with $\Delta_{\rm probe}=205$ 
$\mu$eV, is associated to the quantum ground state transition, occurring at 
$V_G=0.71$ V. 
(c) Extracted bound state energy $E_B$ versus gate $V_G$ and dimensionless
theoretical Kondo temperature $T_K^\mathrm{th}/\Delta$, in comparison to theoretical 
predictions (lines). The dashed vertical line indicates the ground state
transition, found at $T_K^\mathrm{th}/\Delta\approx0.26$. (d) Evolution of the 
conductance peak intensities with $V_G$, showing a kink at the transition,
consistent with a sharp unscreening transition.}

	\label{super}
\end{figure} 

We now turn to the study of the S--QD--S transistor at zero magnetic field.
In presence of superconductivity in both leads, a transport gap of total width 
$2(\Delta+\Delta_{\rm probe}) \approx900 \, \mu$eV opens in the transport map, 
see Fig.~\ref{super}b. The Kondo peak is suppressed and two sharp symmetric 
resonances appear at a certain biasing voltage $V$ so that
$ 0< |eV| -\Delta_{\rm probe} < \Delta$. We take care to differentiate
the gap $\Delta\approx245 \, \mu$eV of the strongly coupled lead, which governs
the physical effects at play, from the gap $\Delta_{\rm probe}\approx205 \,
\mu$eV of the weakly coupled electrode, which offsets essentially the conductance 
onset thresholds by $\pm \Delta_{\rm probe}$. Thermal excitations at 940 mK provide unambiguous evidence of the probe's gap size (see \mbox{Fig.} \ref{temperature}a)

The presence of the YSR states is reflected by extremely sharp subgap resonances 
(\mbox{Fig.} \ref{super}a) at $|eV|=|E_B|+\Delta_{\rm probe}$, that is, when the
probe's chemical potential allows for driving the dot to its excited state by
either adding or removing an electron. The transport mechanisms leading to a
d.c. current here are essentially based on Andreev processes \cite{Ruby2015}. 
From the experimental gate dependence of the bound state spectrum $E_B$, the 
singlet-doublet ground state transition, occurring for $E_B=0$, is readily seen
to occur near $V_G=0.71$ V. Note that
owing to the very sharply defined gap edge of aluminum \cite{vanZanten2015}, but also low experimental
temperatures and careful shielding of the experiment, we can achieve a
spectroscopic resolution down to a FWHM of less than 10 $\mu$eV (\mbox{Fig.} \ref{super}a), way
below previously reported line widths. The latter have indeed been discussed as a
lifetime limiting factor in possible subgap state-based qubits \cite{Pillet2010,
Zazunov2003}.

Combining our knowledge of the superconducting and normal state properties, we 
can now plot the bound state dispersion $E_B$ as a function of gate voltage $V_G$, which 
we express as a function of the dimensionless ratio $T_K^G/\Delta$ (\mbox{Fig.} \ref{super}c). 
We find the transition to the unscreened ground state for the critical value 
$(T_K^\mathrm{G}/\Delta)_c\simeq0.7$, consistent with Ref.~\cite{Buizert2007} or
with the value $(T_K^{FW}/2\Delta)_c\simeq 1.0$ in \mbox{Ref.} \cite{Bauer2013}.      
Theoretical calculations~\cite{Bauer2007} predict a critical value 
$(T_K^\mathrm{th}/\Delta)_c\simeq0.30$, using the scaling formula
$T_K^\mathrm{th}\simeq 0.28\sqrt{U\Gamma}\exp[\pi \epsilon_0
(\epsilon_0+U)/(2\Gamma)]$. While we can rescale our data to $T_K^\mathrm{th}$ (for
instance at the value at the center of the diamond), which gives a reasonable value
$(T_K^\mathrm{th}/\Delta)_c\simeq0.26$, we emphasize that our device is not
strictly in the scaling regime where these predictions apply quantitatively~\cite{Meden2019}.
In addition, a calculation using renormalized ABS theory~\cite{Meng2009}
allows us to obtain a full gate-dispersion of the bound state in good
agreement with the experimental observation, see Fig.~\ref{super}c and Supplemental 
Material file for details.
Furthermore, the intensities of the conductance peaks, which reflect the weigths
carried by bound states, also follow the expected behaviour \cite{Bauer2007} across 
the ground state transition, as shown in Fig.~\ref{super}d.

\begin{figure}[t]
\centering
\includegraphics[width=1.00\columnwidth]{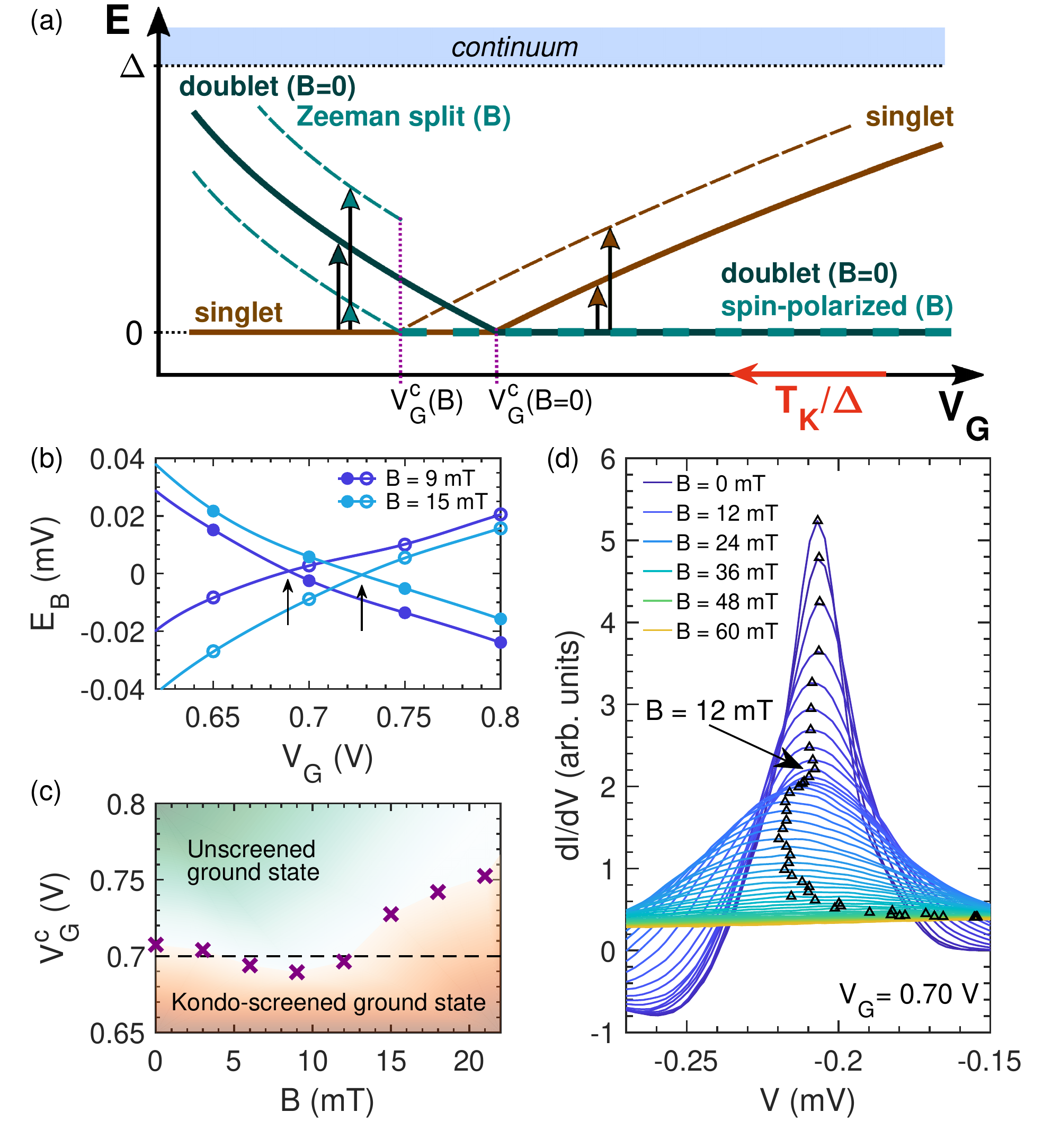}
\caption{(a) Sketch of the level structure of YSR states at zero
magnetic field (continuous lines) and finite (dashed lines) magnetic field, in vicinity
of the ground state transition.
(b)  Zoom on the zero-energy crossings of the bound states dispersion for
two magnetic field values, $B=9$ and 15 mT (symbols). The solid lines are a spline 
interpolation of the data, from the crossing of which the value of the critical gate  
voltage $V_G^c$ is determined (arrows).
(c) Phase diagram of the screening transition in ($V_G$,$B$) space, showing a 
re-entrant phase boundary. 
(d) Spectroscopic analysis of the YSR states as a function of bias and
at a fixed gate voltage of $V_G=0.70$ V, close to the zero field critical point at $V_{\rm G}^{\rm c}= 0.71$ V.
By increasing the magnetic field, a sharp kink is seen indicating the re-entrant phase boundary at $B=12$ mT.}
\label{field}
\end{figure}

Having understood in detail the zero-field properties of the QD--S hybrid, 
we now move to the main result of this work, in which we evidence the
competition of two magnetic effects on the ground state transition of the
quantum dot. A magnetic field $B$ is expected to Zeeman split the two
spin projections of the doublet state by $E_Z=\pm g\mu_B B/2$, with $g$ the
gyromagnetic factor and $\mu_B$ the Bohr magneton. The effect of the Zeeman
splitting on the doublet state has been  observed 
in superconductor - quantum dot junctions formed in semiconducting nanowires, owing to the
large $g\sim20$ in these materials \cite{Lee2014,Jellinggaard2016, Li2017}. In
these works, the sub-gap
resonances are Zeeman split at the singlet ground state phase because two excited
states are accessible. In contrast, when the singlet is the excited state,
no splitting was seen, because the only possible transition (at low temperature $T<E_Z$)
is from the lower energy spin-polarized state to the singlet. 

Beyond the mere spectroscopic effect of the
Zeeman splitting of the doublet excited state at a given dot level depth,
we now consider the magnetic field effect on the
ground state transition itself. Indeed, as one of the spin projections of the
unscreened spin state has a lower energy, the screened ground state phase
space gradually shrinks, which is translated here into a critical value of
$V_G$ moving to lower values. This is sketched in Fig.~\ref{field}a and precisely 
observed in Fig. \ref{field}c, where we plot the critical gate value $V_{\rm G}^{\rm c}$ associated 
to the ground state transition as a function of magnetic field. 
The latter is determined as previously from the kink (crossing)
in the YSR dispersion, for each applied magnetic field as shown in
Fig.~\ref{field}b.
For small fields $B<10$ mT, there is a clear downward trend of $V_{\rm G}^{\rm c}$, indicating 
a Zeeman-driven reduction of the parameter space associated to the singlet ground
state. As the magnetic field is further increased, the reduction of the
superconducting gap starts coming into play, with a quadratic magnetic field
dependence of the gap to lowest order \cite{Anthore2003}. Intuitively, the gradual
weakening of superconductivity favors Kondo screening of the 
spin in the dot, and thereby favors the singlet ground state, enhancing again the
critical $V_{\rm G}^{\rm c}$ (Fig. \ref{field}c).
This re-entrance of the phase-boundary is confirmed when sweeping the magnetic
field at a fixed gate voltage $V_G=0.70$ V (Fig. \ref{field}d).  The transition
of the ground state parity induced near a field of 12 mT is accompanied by an
abrupt change in the YSR spectra, which move to higher energies and acquire a
broader lineshape (see Supplemental Material file for details).

\begin{figure}[t]
\centering
\includegraphics[width=1.0\columnwidth]{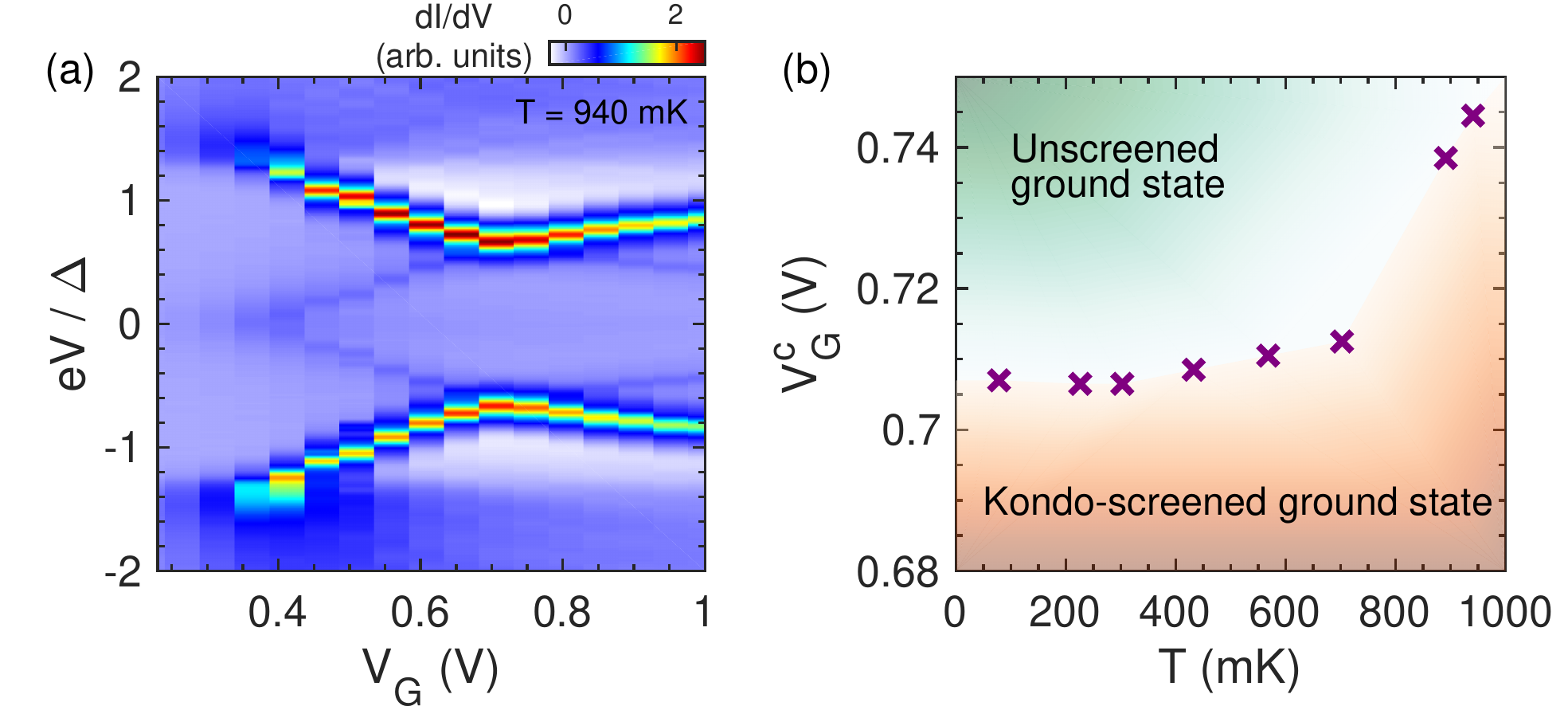}
\caption{(a) Differential conductance of device A in the superconducting
state at the higher temperature $T=940$ mK, displaying the YSR resonances and 
their thermal replicas. 
(b) Gate dependence of the YSR states at 940 mK, as found from the
main conductance resonances (similar data is found from  the thermal replicas).
Owing to the temperature-driven reduction of the superconducting gap $\Delta$, 
the ground state transition has moved to the larger critical gate voltage $V_G^c\approx 0.74$ V
at the largest measured temperature.}
\label{temperature}
\end{figure}

For completeness, we finally focus on the effect of higher temperatures for the
tunnelling spectroscopies as well as the ground state transition.  First, at
higher temperatures, 
the non-zero probability of finding the dot in its excited state
allows for new conductance resonances in tunneling spectroscopies
emerging at $
|eV| = \Delta_{\rm probe} -|E_B|$, which are commonly referred to as {\it thermal
replicas} of the YSR resonances. This is readily seen as a pair of new peaks at
low voltages in Fig.~\ref{temperature}a. The corresponding
values of the bound state energy $E_B$ can now also be deduced from the related supplementary
threshold conditions, in
excellent agreement with the bound state energies deduced from the {\it main}
resonances, leading to the data shown in Fig. \ref{temperature}b. The
singlet-doublet transition can thus be equally observed from the thermal
replicas.
Second, the thermal weakening of the superconducting gap provides another method for 
tuning the singlet-doublet ground state transition. At $T=940$ mK, we indeed find 
that the transition has moved to a larger gate value, about $V_{\rm G}^{\rm c}\simeq 0.74$ V
(or equivalently at a lower $T_K$ than for the base temperature), in agreement with 
expectations. Obviously, no re-entrance is observed in the temperature dependence of the transition.

In conclusion, we have demonstrated a magnetic field tuning of the screening-unscreening 
transition of a quantum dot coupled to superconductors in a transistor geometry.
A novel phase diagram was established, demonstrating that
the magnetic field leads to a re-entrant transition due to the competition
between Zeeman stabilisation of the lowest spin-polarised orbital and weakening
of the superconducting gap. 
A complementary finite temperature phase diagram was drawn, that reflects
the sole thermal weakening on the superconducting gap, while signatures of the 
ground state transition were also observed in thermally excited replicas.
These results demonstrate that quantum dots constitute a rich model system for
the controlled exploration of strong correlations effects in nanostructures. 
Further developments will address the influence 
of the screening-unscreening transition on operational properties of single 
electron turnstiles \cite{vanZanten2016}.

\bigskip

This work was funded by the joint ANR-DFG grant JOSPEC and the Labex LANEF programme. 
Samples were fabricated at the Nanofab facility at Institut N\'eel-CNRS and PTA-CEA. 
We thank D. Basko, J. Pekola, N. Hatter and B.~Heinrich for useful discussions.

\bibliography{./QPT}

\clearpage


\widetext

\begin{center}
\textbf{\large Supplemental Information File: Magnetic-Field-Induced Transition in a Quantum Dot Coupled to a Superconductor}
\end{center}

\makeatletter
\renewcommand{\theequation}{S\arabic{equation}}
\renewcommand{\thefigure}{S\arabic{figure}}

\global\long\def\theequation{S\arabic{equation}}
\global\long\def\thefigure{S\arabic{figure}}
\renewcommand{\thetable}{S\arabic{table}}

\bigskip

This supporting information discusses the sample fabrication process of
electromigrated superconducting quantum dots, the electrical characterization of the 
Kondo-correlated 
quantum dot in both the normal and superconducting state, and presents a second
sample showing similar physics to the one discussed in the main text. Theoretical modeling 
is also developed in detail, including Numerical Renormalization Group (NRG) calculations 
used to extract microscopic parameters of the device, as well as the determination of 
Andreev Bound state dispersions from renormalized perturbation theory.

\bigskip

\section{Sample fabrication}
\label{sec:FAB}
\begin{figure}[ht]
	\includegraphics[width=0.7\textwidth]{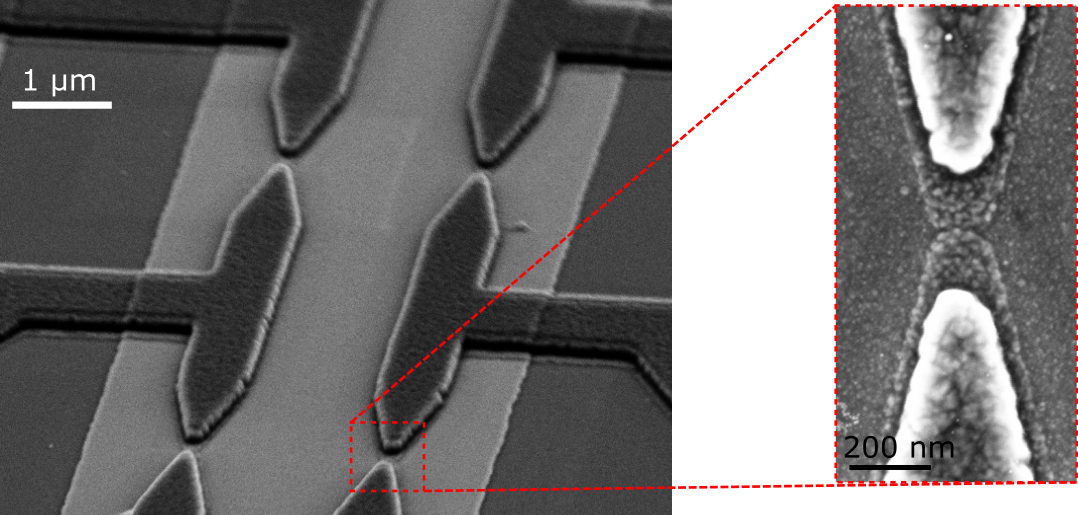}
	\vspace{-0.0cm}
	\caption{Scanning electron microscope images of two \ce{Al} constrictions arrays, fabricated on top of the \ce{ZrO2}-covered gate electrode (left), and a zoom-in at a single constriction, already electromigrated (right).}
	\label{nanofabrication}
\end{figure}

The substrates employed for the fabrication of gated electromigration junctions
are 2-inch intrinsic silicon $<100>$ wafers with a native oxide layer and a
resistivity larger than 8000 $\Omega$ cm. Three steps of lithography, followed
by the corresponding metal deposition and lift-off, are carried out.  First, a
local back-gate is patterned by optical laser lithography using a bi-layer
resist of LOR3A/S1805. After development, a deposition of 3 nm of \ce{Ti}
followed by 30 nm of \ce{Au} is done on the sample with an electron beam
evaporator and put into lift-off solvent. Then the sample surface is cleaned
with a reactive ion etching oxygen plasma of 20 watts power for 5 minutes, and a
\ce{ZrO2} oxide layer of approximately 18 nm is conformally grown by atomic
layer deposition (ALD) technique. In the second lithography step, the bonding
pads and access lines are patterned with a hard-mask UV aligner on a LOR3A/UV3
bi-layer resist. The sample is then metallized with 3 nm of Ti plus 50 nm of Au
and put into lift-off solvent. Electron-beam lithography is employed to pattern
the bow-tie shaped constriction on top of the oxide-covered gate electrode in
the last lithography step. A bi-layer resist of PMMA/MMA AR-P 617.06 and PMMA
3\% is used to create a stable undercut by proximity effect of the beam.
Suspended resist bridges at the narrowest part of the constrictions result from
this undercut. By performing a double angle evaporation of 14 nm of \ce{Al}
fixing the sample at angles of $\pm25^{\circ}$ w.r.t. the axis perpendicular to
the gate electrode, nano-constrictions are grown below each undercut bridge (as
shown in Fig. \ref{nanofabrication}). 

The process finishes with a perpendicular
evaporation of 80 nm of \ce{Al} to decrease the access lines resistance.  Few
hundreds of individual chips are fabricated simultaneously from a 2-inch wafer,
each containing 30 \ce{Al} electromigration constrictions divided in two arrays.
A colloidal toluene suspension of 5 nm diameter \ce{Au} nano-particles
(functionalized with dodecanethiol ligands) is dropcasted several times on the
surface of an individual chip the same day that it is glued and bonded to a
sample-holder and inserted in a dilution cryostat. Electromigration is performed
inside the fridge at a cryogenic temperature of about $4$ K. Junctions
displaying a tunneling current with gate-dependent features are further
investigated at dilution temperatures of $75$ mK (increasing to 100 mK for magnetic field experiments). Around 100 aluminum
constrictions have been successfully electromigrated in the work of the present
article. Several of these breaking tunnel junctions presented a gate dependent
current, but only 4 displayed an identifiable diamond-like structure in the
stability diagram and sub-gap current that indicated the presence of bound
states. Out of 4, only 2 (samples A and B) had a gate dependence stable enough
to acquire systematic successive measurements.

\section{Extracting experimental parameters from NRG}
\label{sec:NRG}

Strongly correlated quantum dots are generically described by a single-level 
impurity model involving two electrodes (left/right, or drain/source), 
according to the Anderson Hamiltonian:
\begin{equation}
H = \sum_{\sigma}\epsilon_{0} d_{\sigma}^{\dagger}d_{\sigma}
+U d_\uparrow^\dagger d_{\uparrow} d_\downarrow^\dagger d_{\downarrow}
+
\sum_{k\alpha\sigma}\epsilon_{k}c_{k\alpha\sigma}^{\dagger}c_{k\alpha\sigma}
+
\sum_{k\alpha\sigma}t_{\alpha}(c_{k\alpha\sigma}^{\dagger}d_{\sigma} + \mathrm{H.c.}),
\label{S:Anderson}
\end{equation}
assuming that confinement is strong enough to disregard fully occupied or empty
orbitals.
Here $d^\dagger_\sigma$ creates an electron on the dot with spin 
$\sigma=(\uparrow,\downarrow)$, and $c^\dagger_{k\alpha\sigma}$ an electron in the lead 
$\alpha=(L,R)$ with spin $\sigma$ and momentum $k$. The parameters of the model in Eq.~(\ref{S:Anderson}) 
are the single-particle energy level $\epsilon_0$ of the quantum dot (relative to
the equilibrium Fermi level of the leads, taken as energy reference), the local Coulomb 
repulsion $U$ on the dot, the kinetic energy $\epsilon_{k\sigma}$ of the lead electrons, and
the tunneling amplitude $t_{\alpha}$ from the dot to electrode $\alpha$. Taking
into account a generic asymmetry between the coupling to each electrode, we
define the respective hybridizations as $\Gamma = \pi \rho_0 (t_L)^2$ and
$\gamma = \pi \rho_0 (t_R)^2$, with $\rho_0$ the electronic density of states at the Fermi 
level (assumed identical in both electrodes, without loss of generality).
Typically, in a tunneling geometry relevant for the present experiment, one has
$\Gamma\gg\gamma$. In addition, a voltage bias $V=\mu_L-\mu_R$, originating from an imbalance 
of the chemical potentials $\mu_\alpha$ in each lead, allows to drive a current through the 
nanostructure, and forms the basis of conductance measurements in electronic quantum dot devices.

We define a bonding orbital $c^\dagger_{k\sigma} = \sum_\alpha
(t_\alpha/\sqrt{t_{L}^2+t_{R}^2})
c_{k\alpha\sigma}^{\dagger}$ and its corresponding orthonormal antibonding
orbital, so that the antibonding orbital decouples from the
tunneling process via the dot level. 
The total hybridization of the $d$-level is thus $\Gamma+\gamma\simeq\Gamma$,
since $\Gamma\gg\gamma$ in a tunneling geometry.
This allows to use a single channel Anderson model in equilibrium, which simplifies the 
numerical simulations.
A Kondo resonance in the dot density of states $\rho(\omega)$ develops provided 
$U\gtrsim\pi\Gamma$ in the local moment regime $|\epsilon_0 + U/2| \lesssim U/2$ where 
charge dynamics is frozen by Coulomb blockade, but spin fluctuations associated to the dot orbital
assist transport through the nanostructure.
One defines also the mixed valence regime between the empty and single occupied
quantum dot as $|\epsilon_{0}|\lesssim \Gamma$.

\begin{figure}[ht!]
\includegraphics[width=0.48\textwidth]{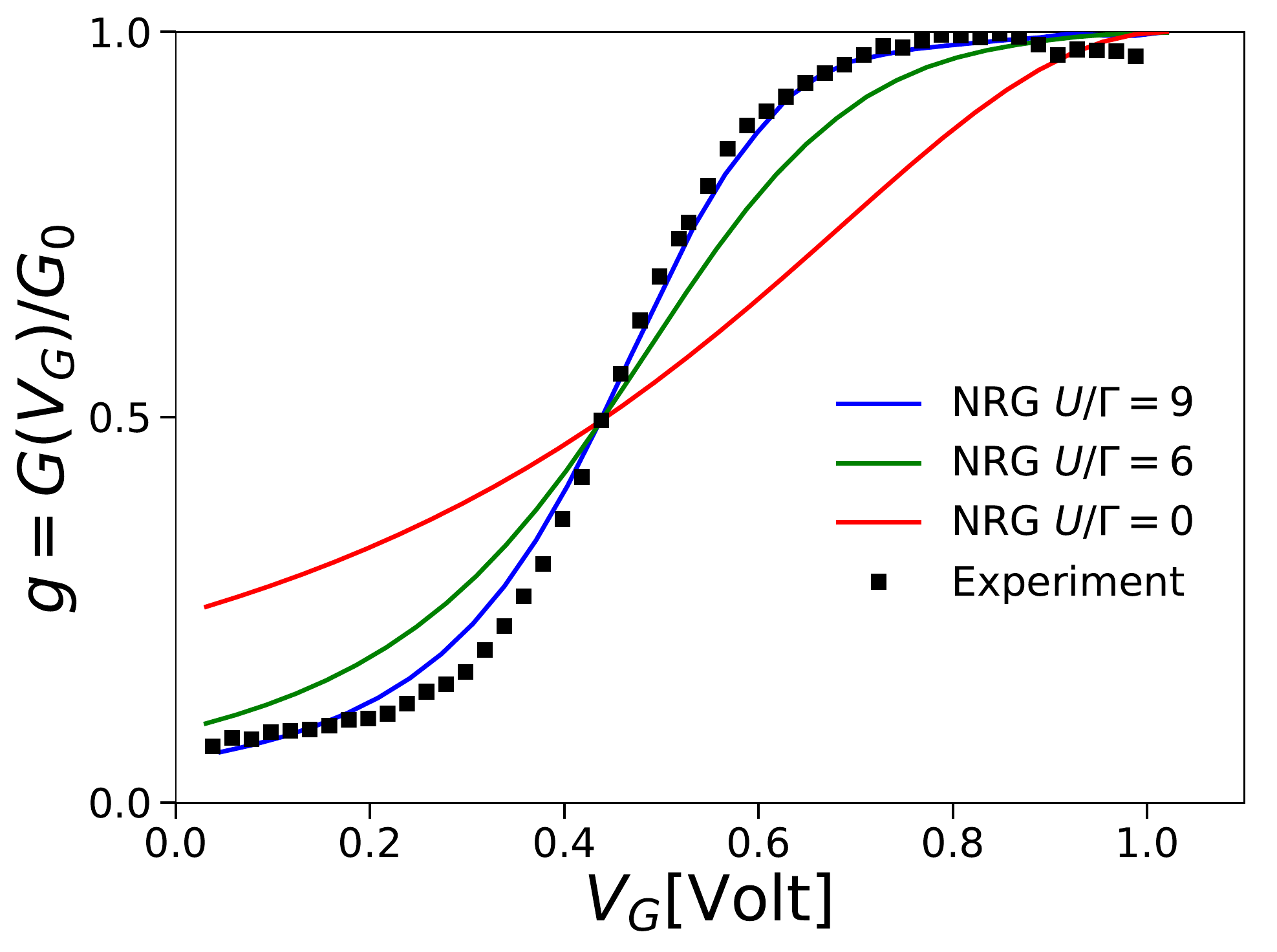}
\vspace{-0.0cm}
\caption{
Rescaled experimental conductance $g(V_G)=G(V_G)/G_0$ (black dots) as a function of gate voltage 
$V_G$ in the experiment. The NRG results (colored lines), shown here for three values of $U/\Gamma=0,6,9$, 
are shifted to match the center of the Coulomb diamond location, and are rescaled horizontally so 
that all NRG curves intercept the experimental data set at the diamond edge
(midpoint $g=1/2$), allowing to test the scaling form of $g(V_G)$. The optimal value $U/\Gamma=9$ 
is inferred from this comparison.}
\label{NRG}
\end{figure}

The most crucial microscopic parameter of Hamiltonian in Eq.~(\ref{S:Anderson}) to be
extracted from the experiment is the ratio $U/\Gamma$, which sets the
strength of electronic correlations. Standard Coulomb diamond spectroscopy
cannot be used in the regime where Kondo correlations are fully developed
due to many-body modification of the excitations' linewidth. A more reliable
method is to use the gate dependence of the zero-bias conductance, which is
readily obtained from linear response theory:
\begin{eqnarray}
\label{S:linG}
G(V_G) & = &
\frac{2 e^2}{h} \frac{4\pi \Gamma \gamma}{\Gamma+\gamma} 
\rho(\omega=0,\epsilon_0) \simeq
\frac{2 e^2}{h} \frac{4\pi \gamma}{\Gamma} 
\rho(\omega=0,\epsilon_0)\\
\label{eps0}
\epsilon_0 & = & -U/2 + \alpha_G ( V_G-V_G^\mathrm{center})
\end{eqnarray}
with $\rho(\omega,\epsilon_0)$ the zero-temperature finite frequency density of states 
of the impurity level, assuming that the base temperature is well below the Kondo scale.  
We model in Eq.~(\ref{eps0}) a linear relation between the level position 
$\epsilon_0$ and the applied gate voltage $V_G$, with
$\alpha_G$ the gate lever arm, and $V_G^\mathrm{center}$ the offset voltage corresponding 
to the center of the Coulomb diamond in the experiment, so that
$V_G=V_G^\mathrm{center}$ 
implies $\epsilon_0=-U/2$ (particle-hole symmetric point).
The maximum value for $G(V_G)$ allowed by Friedel sum rule
is $G_0 = (2 e^2/h) 4 \Gamma \gamma/(\Gamma+\gamma)^2$, a value that is
attained at the center of the odd charge Coulomb diamond (namely $\epsilon_0 =-U/2$)
at zero temperature. We define a rescaled dimensionless conductance $g(V_G) =
G(V_G)/G_0$ that crosses over from 0 (in the empty dot regime
$\epsilon_0>0$) to 1 (at the center of the odd charge diamond $\epsilon_0 = -U/2$).
The lineshape of $g(V_G)$ is interesting to estimate correlations because it
forms a flat plateau of width proportional to $U$, and displays a crossover on a 
scale of the order of $\Gamma+\gamma$.
In addition, $g(V_G)$ is weakly sensitive to the electronic bandwidth, provided
that it is larger than hybridization (not shown).
Since $g(V_G)$ is mostly controlled by the single scale
$U/\Gamma$, this ratio can be extracted precisely from a scaling analysis.
A comparison of the experimental measurement to the NRG data is shown in 
Fig.~\ref{NRG}. 
We have chosen here three values of $U/\Gamma=0,6,9$, and fixed the ratio 
$\Gamma/D=0.3$, where $D$ is the half-bandwidth of the electrodes.
The horizontal axis corresponds to the experimental voltage $V_G$, and 
a small charge offset artifact was removed from the experimental data in order to
produce a smooth curve. This results in a small offset of the 
center of the Coulomb diamond expected near $V_G^\mathrm{center}=1.1$V.
Regarding the rescaling of the NRG data, we apply a similar centering to the
Coulomb diamond center, $\delta \epsilon_0 = \epsilon_0 + U/2$. Then we convert 
the energy level to voltage from the multiplicative form 
$\delta V_G = \delta \epsilon_0 / (-\alpha_G)$. For each NRG curve, we choose 
$\alpha_G$ so that the normalized NRG conductance $g(V_G)$ crosses 
the experimental points at the mid point $g=1/2$. This allow to test the
scaling form of $g(V_G)$ for various values of the parameters. We find that
our experimental data are best fitted for the value $U/\Gamma=9$. One
notices a slight decrease of $g(V_G)$, below the unitary limit, in the experimental data 
for the largest $V_G$ values. Such a thermal effect is expected from the finite temperature (100 mK) at
which the measurements were done.

\section{Characterization of sample A}
\label{sec:SAMPLE_A}
The nanometric crack appearing in electromigrated constrictions after the breaking procedure has normally an irregular shape and the distance between the two resulting leads is variable along the constriction section, often below 1 nm at certain spots. Usually, a tunneling current directly from one lead to the other is measurable, bypassing any possible quantum dot and originating a conductance background. We start the characterization of the sample by measuring the linear shunt resistance of the junction, $R_S$, away from the degeneracy point in the even occupation side. We found $R_S \approx 20$ M$\Omega$ at $V_G = -0.2$ V in sample A.

\subsection{Coulomb blockade analysis}

\begin{figure}[t!]
	\includegraphics[width=1.0\textwidth]{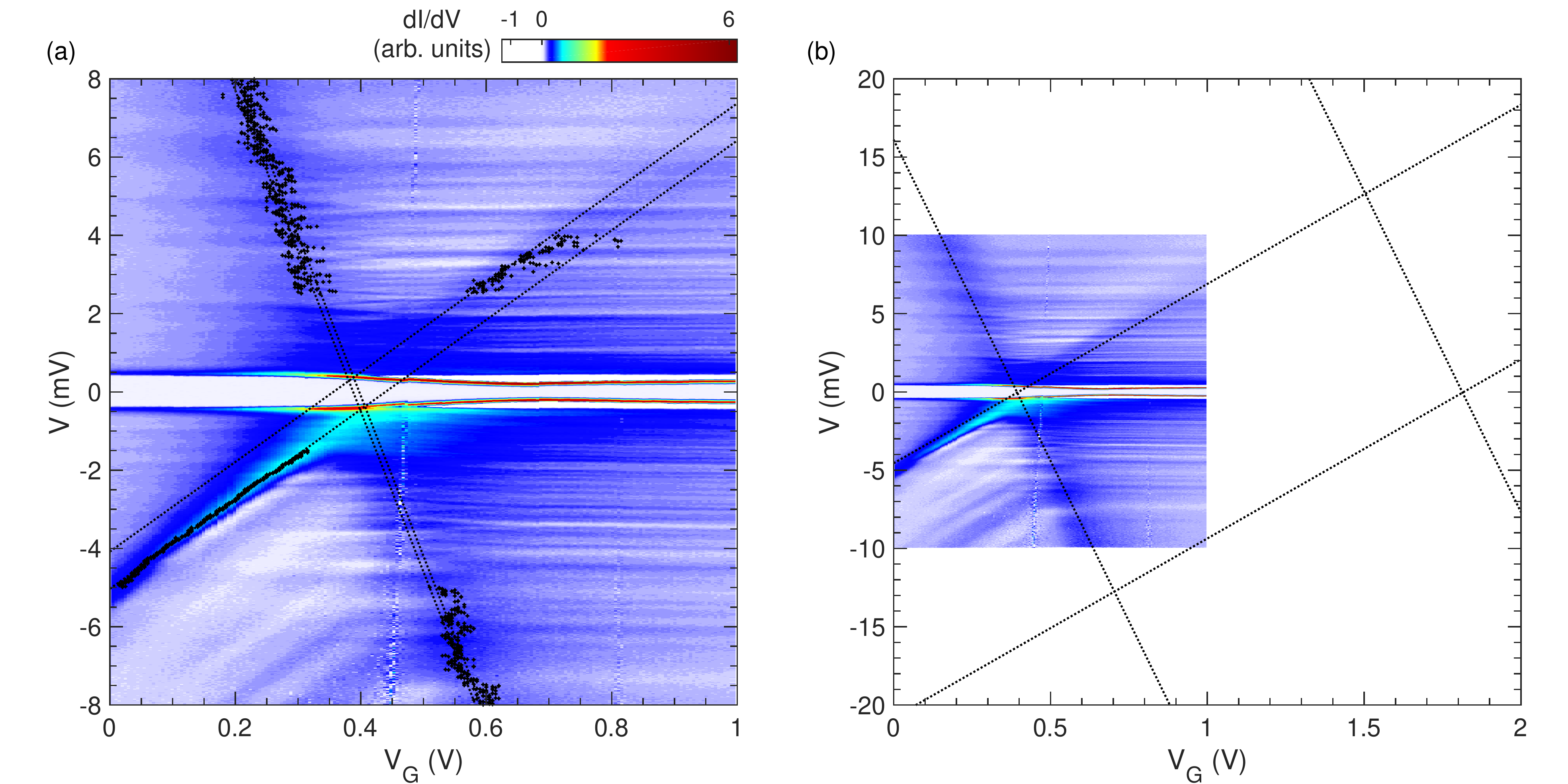}
	\caption{(a) Superconducting state $dI/dV$ differential conductance
stability diagram of sample A. Linear fits (dashed lines) of the extracted
conductance maxima (black dots) have been done to infer the four diamond edges.
A certain displacement between linear fits with the same slope is introduced by
the opening of a spectroscopic gap at zero $V$ in the superconducting state (b)
By plotting the linear fits obtained in (a) at the next degeneracy point ($V_G^1
\approx 1.8$ V), extrapolated from the parabolic fit in Fig. \ref{fig:TKfit}c,
we obtain a full picture of the stability diagram allowing to extract an
approximated value of $U \approx 12.7$ meV from the crossing of the linear fits
at finite $V$.
	}
	\label{fig:Cap_coup}
\end{figure}

The well-known diamond like structure of blocked current regions appearing in
the stability diagram of gated-quantum dot junctions become blurred in the
presence of higher order transport processes (like the spin-flip Kondo
resonance), enhanced due to a high relative tunnel coupling energy scale
(${\Gamma} \gg k_BT$). An appropriate extraction of the diamond features is
crucial to determine quantitatively the relevant parameters of the quantum dot
junction, such as its degeneracy point position ($V_G^0$) in the gate voltage
dimension, its capacitive coupling to the source, drain and gate electrodes
($C_S, C_D, C_G$ respectively) and its charging energy $U$. The sharp density of
states of the superconducting \ce{Al} leads promotes the diamond edges
differential conductance, allowing us to perform linear fits of the
corresponding conductance maxima, as shown in Fig. \ref{fig:Cap_coup}a. Adapting
Coulomb blockade standard analysis methods \cite{thijssen2008charge} to our
particular case of superconducting leads, the degeneracy point is found to be at
$V_G^0 \approx 0.40$ V from the middle point between the crossing of the linear
fits at $V = 0$, in excellent correspondence with the mid point $g = 1/2$ at
which the measured zero-bias conductance of the Kondo resonance decreases by
half, providing a criterium to select the numerical values of $\alpha_G$ for the
comparison of the NRG simulations in Fig. \ref{NRG} with the experimental data.
The experimental gate lever arm $\alpha_G$ and the source--drain capacitance
asymmetry have been determined from the extracted positive ($\beta = 11.5$ meV/V) 
and negative ($\beta^{\prime} = -41$ meV/V) diamond edges slopes using
the following relations:
\begin{equation}
\alpha_G = \frac{C_G}{C_S + C_D + C_G} = \frac{\beta\beta^{\prime}}{\beta + \beta^{\prime}} \approx 9 \,\text{meV/V},
\label{Gate_coup}
\end{equation}
\begin{equation}
\frac{C_D}{C_S} = \frac{\beta^{\prime}}{\beta(1-\beta)} \approx 3.5\,.
\label{SD_coup}
\end{equation}
The high source--drain capacitance asymmetry derived for sample A is reflected
in a few orders of magnitude difference in the tunnel couplings (calculated
later in this section), leading to a probing configuration of a strongly coupled
system, commonly reached in scanning electron microscopy experiments. The small
$\alpha_G$ value, typical from single molecule/nano-particle transistors, does
not allow to observe the next degeneracy point ($V_G^1$) of the quantum dot
junction. However, as explained in the following, we performed a parabolic fit
of the Kondo temperature gate dependence that indicates the position of the
Coulomb diamond center $V_G^\mathrm{center}$ to be close to $1.1$ V. An extrapolation of
the next degeneracy point can be done so that $V_G^1 = V_G^\mathrm{center} +
(V_G^\mathrm{center}-V_G^0) \approx 1.8$ V (see Fig. \ref{fig:Cap_coup}b), leading to a
charging energy of $U \approx 12.7$ meV, in good agreement with the value found
for the experimental gate coupling of around 0.9 \%.

\subsection{Alternative Kondo temperature extraction methods}

\begin{figure}[htb!]
	\includegraphics[width=1.0\textwidth]{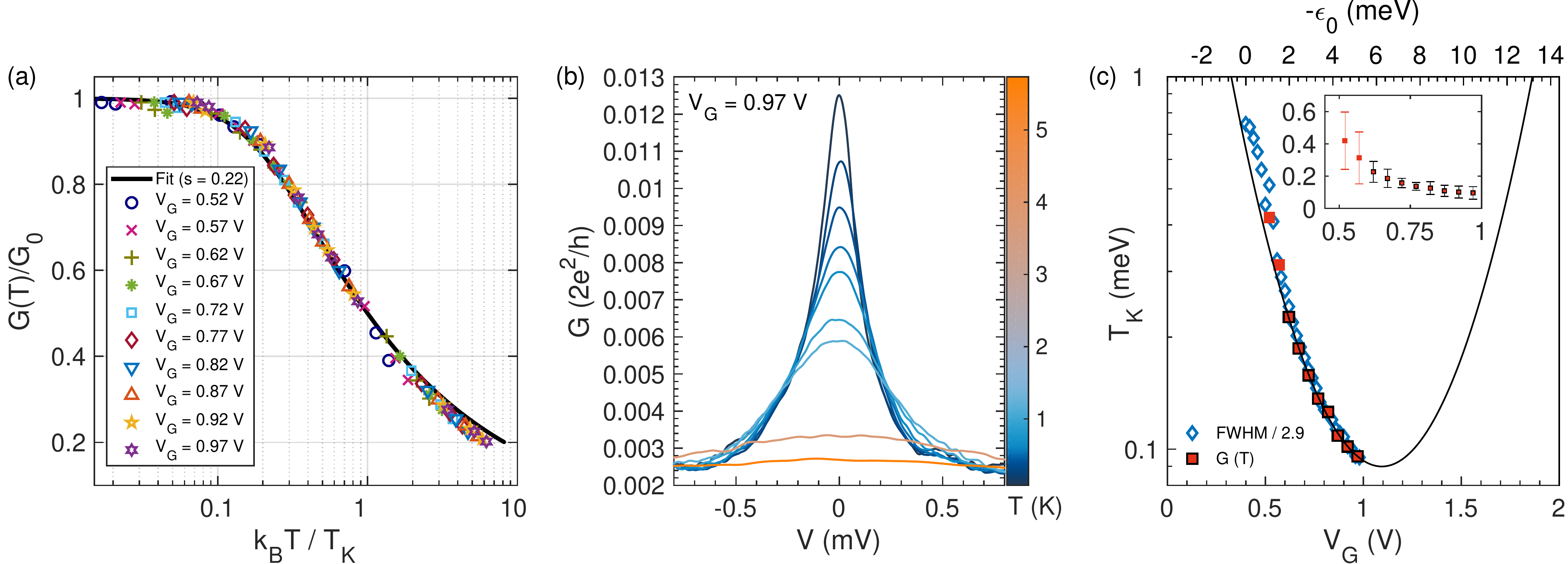}
	\caption{(a) The temperature dependence of the zero-bias linear
conductance shows an universal behavior for different $V_G$ ($\epsilon_{0}$),
well described by Eq. (\ref{eqn:GGfit}). The $T_K$ normalizing value at each
$V_G$ have been found using a least squares fitting method. (b) Differential
conductance spectroscopies of the Kondo resonance at different temperatures for
$V_G = 0.97$ V. (c) The gate dependence displayed by the extracted $T_K$ values
in (a) is identical to the low temperature FWHM of the resonance (scaled by a factor of 1/2.9). A qualitative parabolic fit according to Eq. (\ref{eqn:TKdef})
has been done for $G(T)$ data points away from the mixed valence regime (framed
squares). The inset shows the same $T_K$ values represented in linear scale,
including error bars showing the standard deviation of the least squares fit.
	}
	\label{fig:TKfit}
\end{figure}

Different criteria can be found in the literature for the experimental
extraction of the Kondo energy scale ($T_K$) in quantum dot junctions. The lack
of a general $T_K$ definition valid for all regimes of relative energy scales
complicates the fitting and comparison of results for samples involving
different ratios between $U$, $\Gamma$ and $\epsilon_{0}$. For quantum dot
junctions identified to be in the Kondo regime ($U\gg\Gamma$), a possible
zero-temperature definition is $T_K = 0.28 \sqrt{\Gamma U}
e^{\frac{-\pi U}{8\Gamma}}$, valid only at the center of the odd
diamond ($\epsilon_0 = -U/2$). Although the $T_K$ dependence remains the same \cite{choi2004kondo}, in a generic electrostatic configuration within the Coulomb diamond (namely $0>\epsilon_0>-U$) the proportionality factor is not well-defined:
\begin{equation}
T_K \propto \sqrt{\Gamma U}
e^{\frac{\pi\epsilon_0(\epsilon_0 + U)}{2\Gamma U}}.
\label{eqn:TKdef}
\end{equation}
Experimentally, $T_K$ is often extracted from the \textit{full width at half
maximum} (FWHM) of the Kondo zero-bias peak, with a variable multiplicative
factor
\cite{buitelaar2002quantum,cronenwett1998tunable,Franke2011,Maurand2012,Bauer2013,lee2017scaling}
generally equal to 1/2. However, thermal broadening and reduction of the
resonance, conductance backgrounds and slight shifts in the relative energy
scales can induce a non-negligible mismatch between different experiments in
Kondo-regime systems. Alternatively, an empirical expression deducted from the
NRG model was proposed by Goldhaber-Gordon et. al. \cite{goldhaber1998kondo}, as
a rigorous method for extracting $T_K$, based on the universal dependence of the
relative resonance maxima conductance with temperature:
\begin{equation}
G(T) = G_0\left(\frac{T^2}{T_K^2}(2^{1/s}-1)+1\right)^s + G_B,
\label{eqn:GGfit}
\end{equation}
where $G_0+G_B$ represents the zero bias conductance at zero temperature (as
explained in the first section \ref{sec:NRG}), $G_B$ represents the conductance
background of the junction (typically $G_0\gg G_B)$, and $s$ is a fitting parameter 
found to be around $s=0.22$ for magnetic Kondo impurities with spin 1/2. In symmetrically coupled quantum dot junctions in the Kondo regime measured at zero temperature, $G(T\rightarrow0) \approx G_0$ is expected to reach the quantum of conductance
$2e^2/h$ around the center of the odd diamond ($-U<\epsilon_0<0$). In contrast, in an asymmetric case where one lead has a reduced coupling down to the
so-called sequential regime ($\Gamma\gg k_BT\gtrsim\gamma$), the absence of
Kondo correlations with this weakly coupled lead ($\gamma$) modulates the peak 
intensity of the Kondo resonance (only present at the strongly coupled lead), 
leading to a strong reduction w.r.t. to the unitary value:
\begin{equation}
G_0 \simeq \frac{2e^2}{h} \frac{4\gamma}{\Gamma^2}.
\label{eqn:NormG}
\end{equation}
In our sample configuration, the presence of a gate electrode allows us to
modify the quantum dot chemical potential ($\epsilon_0 = \alpha_G V_G$ up to
an arbitrary offset at $V^{center}_G$), leading to a gate-tunable $T_K$ parameter. We have measured 
and performed a least squares fit of $G(T)$ at different temperatures of sample A with
Goldhaber-Gordon's Eq. (\ref{eqn:GGfit}), for each accessible value of $V_G$
(Fig. \ref{fig:TKfit}a). 

The resulting values of $T_K$ were found by setting
$G_B = 0$ (since the differential conductance away from the degeneracy point
($V_G = -0.2$ V) decreases below $5\%$ of $G_0$), and imposing an extrapolated
conductance at zero temperature of $G_0 \approx 0.013$ (Fig. \ref{fig:TKfit}b).
Additionally, the value of $G_0$ can be plugged in Eq. (\ref{eqn:NormG}), leading
to a tunnel coupling asymmetry of $\Gamma\approx300\gamma$. Combining the
conclusions obtained from NRG simulations for the relative energy scales value of
$U/\Gamma = 9$ with the extraction of $U \approx 12.7$ meV from the Coulomb
blockade analysis, we derive the tunnel couplings to be $\Gamma \approx 1.4$ meV
and $\gamma \approx 5$ $\mu$eV.

The Kondo resonance FWHM, measured as a function of bias voltage and at a base temperature of $T = 100$ mK ($k_BT = 8.6$ $\mu$eV), displays an identical gate-driven renormalization, albeit for a scaling factor of 1/2.9 w.r.t. to the previous temperature-based extraction (Fig. \ref{fig:TKfit}c). The natural logarithm of the extracted $T_K$ displays a clear parabolic dependence that can be fitted with the previously given $T_K$ definition in Eq. (\ref{eqn:TKdef}), using the extracted value for $\alpha_G$ to
determine $\epsilon_0 (V_G)$. This quadratic fit provides an alternative way of
extracting the energy scales $U$ and $\Gamma$, obtaining $U=$ 12.5 meV and
$\Gamma=$ 2.5 meV respectively. Although the resulting $U$ is extremely close to the value inferred from the Coulomb Blockade analysis (12.7 meV), this sample is not deep
enough in the Kondo regime, as derived from the NRG simulations ($U = 9\Gamma$),
so that the quadratic fit can be only taken into account qualitatively. However, the 
overall tendency of $T_K$ variation with $V_G$ is well described by the fitting
parabola, leading us to designate the center of the blocked diamond
($V_G^\mathrm{center}$) employed in the Coulomb blockade analysis from its minimum at
about 1.1 V. As $V_G$ approaches $V_G^0$, the system enters into the mixed
valence regime ($|\epsilon_0|\lesssim\Gamma$), where Eq. (\ref{eqn:TKdef}) is not
valid anymore and the tendency of the $T_K$ gate dependence changes. Although the least
squares fitting method still provides a normalizing $T_K$ value for voltages of
$V_G =$ 0.52 V and 0.57 V, and the temperature dependence of the experimental data follows the
universal behavior in rather good agreement, we observe in the inset of Fig.
\ref{fig:TKfit}c (where $T_K$ is represented linear scale) that the standard
deviation suffers a sudden increase, reaching values around 0.2 meV. Therefore,
the corresponding $T_K$ values (shown in Fig. \ref{fig:TKfit}c as unframed
squares) cannot be highly trusted, and have not been taken into account for fitting the parabola.

\subsection{Magnetic field behavior in the superconducting state}

\begin{figure}[b!]
	\hspace{1cm}
	\includegraphics[width=1.0\textwidth]{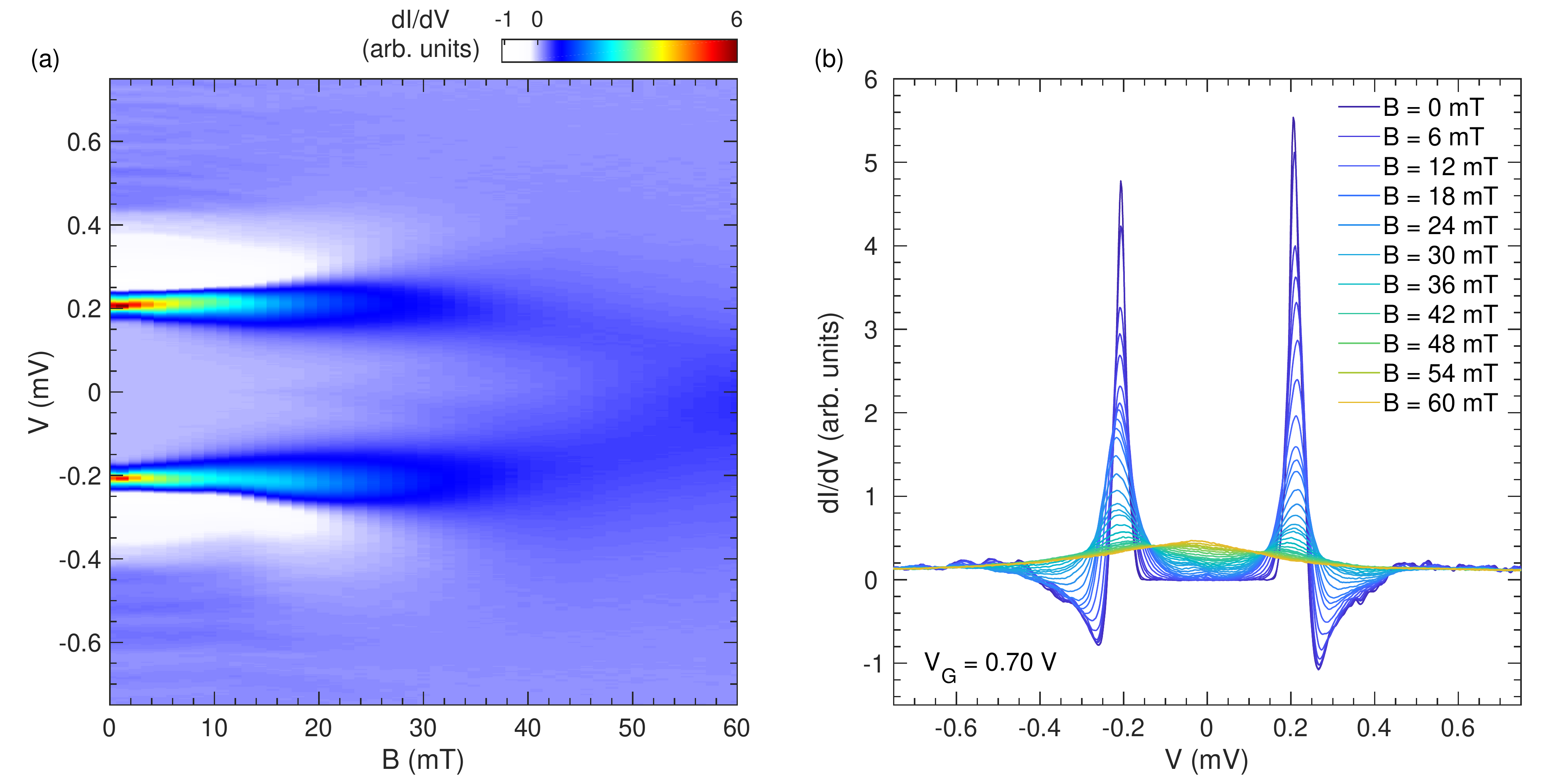}
	\caption{(a) dI/dV differential conductance map for different values of the magnetic field at $V_G = 0.7$ V. A gradual transition from the subgap YSR states to the zero-bias Kondo resonance takes place in the spectra as magnetic field quenches superconductivity in the leads. (b) Line cuts of the map showing the spectrum evolution for increasing magnetic fields.
	}
	\label{fig:BMAP}
\end{figure}

When turning the strongly coupled lead into the superconducting state, Yu-Shiba-Rusinov (YSR) subgap excitations emerge from the competition between the superconducting electron pairing ($\Delta$) and the Kondo correlations induced by the oddly-occupied quantum dot \cite{Franke2011}, probed at different $V_G$ by the weakly coupled lead, which is also in superconducting state. As discussed in the main article, the crossing of the YSR resonances at zero energy indicates a phase transition happening at $T_K/\Delta \approx 0.7$ (at a critical gate voltage value of $V_G^c = 0.71$ V), where both of the ground states at each side of the critical point (the Kondo singlet at the 0-phase and the unscreened doublet at the $\pi$-phase) undergo an inversion, becoming the excited state at the contrary side. 

By applying an external magnetic field ($B$) perpendicular to the sample surface, it is possible to quench superconductivity, turning the leads into
the normal state, suppressing the superconducting energy scale $\Delta$.
In Fig. \ref{fig:BMAP} we can observe how the spectroscopic gap (with a total
range of $V \approx \pm 0.45$ mV at $B = 0$ mT) decreases as $B$ is increased,
and has completely disappeared at $B = 50$ mT. Accordingly, the YSR peaks
decrease their intensity and finally merge into the Kondo resonance at zero
bias, much broader and less intense. Tuning the gate to 0.70 V, close to the critical point
(at $V_G^c$ = 0.71 V), and measuring the spectrum at different $B$, a gentle
oscillation in the energy of the negative bias YSR peak can be observed, as
shown in the zoomed-in graph of Fig. 3d in the main article. In order to
further investigate this phenomenon, we measured whole differential conductance
maps (analogous to that one shown in Fig. \ref{fig:Cap_coup}a) at different
fixed values of $B$. The critical gate position $V_G^c$ displays an apparent
oscillation, reaching a minimum value at 9 mT. This downwards tendency,
enlarging the $\pi$-phase effective area is a consequence of the Zeeman
splitting of the excited state at the 0-phase, that displaces the phase boundary
position. For increasing $B$ above 9 mT, the quenching of superconductivity in the
lead gradually reduces the effective $\Delta$, favoring the dominance of the
0-phase, i.e. the formation of the Kondo singlet.

\section{Andreev bound states dispersion from renormalized perturbation theory}
\label{sec:ABS}
In presence of superconductivity, we supplement the Anderson impurity 
model~(\ref{S:Anderson}) with a pairing contribution in the electrodes: 
$H_\mathrm{pair} = 
- \sum_{k\alpha} \Delta_{\alpha}^{} \, c_{k,\uparrow,\alpha}^{\dagger} 
c_{-k,\downarrow,\alpha}^{\dagger} + \mathrm{h.c.} $, with $\Delta_{L/R}$ 
the superconducting gap in the left and right lead respectively. 
In order to gain some insight, we discuss first briefly the large gap solution of the
superconducting Anderson model~\cite{H_eff_original_affleck,
Delta_inf_assym_hewson,Bauer2013,vecino_yeyati}, and make as previously
the assumption of asymmetric tunneling, $\Gamma\gg\gamma$.
The bare Green's function of the dot indeed becomes purely
local for $\Delta_{L/R}\to\infty$ (we use here Nambu notation):
\begin{align}
\label{NonIntG}
G_0(i\omega) = 
\begin{pmatrix}  i\omega - B - \epsilon_0  & \Gamma \\  \Gamma  & 
i\omega- B + \epsilon_0  \end{pmatrix}^{-1},
\end{align}
which derives from an effective local Hamiltonian
\begin{equation}
H_{\rm{\rm eff}}^0 =  \Psi^\dagger \begin{pmatrix} B + \epsilon_0 & -\Gamma \\
-\Gamma & B - \epsilon_0 \end{pmatrix} \Psi,
\label{ALHamNI}
\end{equation}
to which one can add the interaction term controlled by the Coulomb repulsion $U$.
Diagonalizing the full local Hamiltonian using a Bogoliubov basis transformation
provides four discrete Andreev eigenstates with the eigenvalues:
\begin{equation}
E_\sigma = \sigma B, \qquad E_{\pm} = U/2 \pm 
\sqrt{\xi_0^2  + \Gamma^2},
\label{ABSspectrum}
\end{equation}
where $E_\sigma$ correspond to spin polarized singly occupied states, while $E_\pm$
are proximity induced superpositions of the empty and doubly occupied states
(BCS-like states). We have noted $\xi_0=\epsilon_0+U/2$, the level position
relative to the particle-hole symmetric point.
A phase transition is readily seen to occur (for $B>0$) when 
$E_\downarrow = E_-$, which reads explicitly:
\begin{equation}
(U+2B)^2 =\; 4 \Gamma^2 + 4\xi_0^2.
\label{PTCond}
\end{equation}
Thus, in absence of Kondo correlations, $U$ and $B$ play a similar role in 
determining the phase boundary, as both tend to induce a transition to a
spin-polarized state, the $\pi$-phase. 
In most practical situations, the gap is not the largest scale in the problem,
and the approximate large gap Andreev spectrum~(\ref{ABSspectrum}) misses 
several physical effects, such as level repulsion at the gap edge, and the competition 
with the Kondo screening process. Both mechanisms are readily accounted for by
a renormalized perturbative approach around the large gap limit, that we
briefly introduce here~\cite{Meng2009,Wentzell}. We first define the bare (in the large gap
limit) Andreev excitation energy $a_\sigma^0 = E_- - E_{\sigma}$.
The full transition excitation spectrum reads within renormalized perturbation theory:
\begin{align}
\label{eq:a}
a_\sigma  = & a_\sigma^0 + \delta a_\sigma =
a_\sigma^0 -\frac{\Gamma}{\pi}\int_{0}^{D}\!\!\!\! d\epsilon \, 
\Bigg[\sum_{\sigma'}\frac{1}{E-a_{\sigma'}^0-\Theta[-\delta a_{\sigma'}]
\delta a_{\sigma'}} -\frac{1}{E+b_\sigma^0}
-\frac{1}{E+a_\sigma^0}\nonumber\\
&+\frac{2\Delta}{E} \frac{\Gamma_D}{\Gamma_D^2+\xi^2}
\Bigg(\sum_{\sigma'}\frac{1}{E-a_{\sigma'}^0-\Theta[-\delta a_{\sigma'}]\delta a_{\sigma'}}
-\frac{1}{E+b_\sigma^0}
 +\frac{1}{E+a_\sigma^0+\Theta[\delta a_{\sigma'}]\delta
a_{\sigma'}}\Bigg)\Bigg] + \frac{2\Gamma_D^2}{\Gamma_D^2+\xi^2},
\end{align}
with the quasiparticle energy $E = \sqrt{\epsilon^2+{\Delta}^2}$, and $\Theta$
the Heaviside function.
Due to the finite bandwidth of the leads, one introduces a generalized hybridization
parameter $\Gamma_D = \frac{2}{\pi} \arctan(D/\Delta) \Gamma$.
Since we assumed a tunneling geometry, $\Gamma\gg\gamma$, only the single gap $\Delta_L$
of the most coupled electrode, denoted $\Delta$, appears above. Solving for $\delta
a_\sigma$ in Eq.~(\ref{eq:a}) provides the wanted Andreev dispersion, allowing
to locate the quantum phase transition when $a_\sigma=0$ as a function of
the various parameters of the system.

\section{Characterization of sample B}
\label{sec:SAMPLE_B}
\begin{figure}[b!]
	\includegraphics[width=0.9\textwidth]{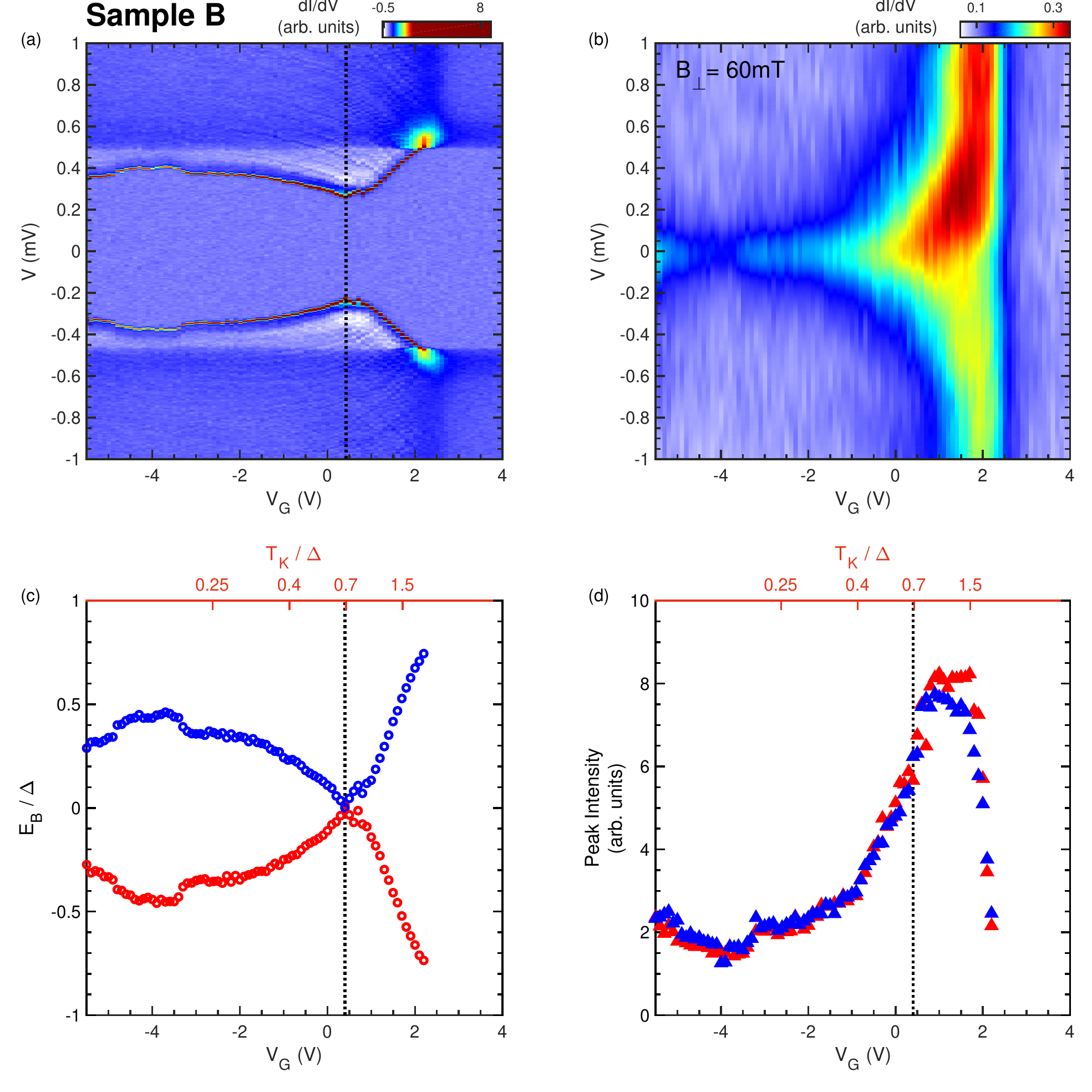}
	\caption{(a) Superconducting
state $dI/dV$ differential conductance map of sample B. The minimal spacing of
YSR spectral peaks happens at $V_G \approx 0.4$, signaling the position of the
phase transition, indicated by a vertical dashed line.
(b) Normal state $dI/dV$ differential conductance map of sample
B. The degeneracy point is located around $V_G \approx 2$ V. 
(c) YSR states' energy dispersion $E_B$ plotted versus $V_G$ and the corresponding 
Kondo energy scale $T_K$ axes. The occurrence of the phase transition is observed 
at $T_K/\Delta \approx 0.7$ (d) Dependence of the YSR states' peak intensities with 
$V_G$ and $T_K$.
}
	\label{fig:sampleB}
\end{figure} 
Following the same headlines as in sample A, the characterization of sample B
starts with the measurement of a linear shunt resistance of 15
M$\Omega$ at $V_G = 4$ V, deep in the even diamond. Values for the
superconducting pairing energy of the probing ($\Delta_{probe} \approx 240$
$\mu$eV) and the strongly coupled lead ($\Delta \approx 330$ $\mu$eV) have been
extracted from the differential conductance map in the S state (Fig.
\ref{fig:sampleB}a) following the same criteria as in sample A. Such an increase
of the superconducting gap size in thin aluminum films with respect to the usual
bulk value has been previously reported and explained as a consequence of confinement
effects and granularity,
\cite{ferguson2007energy,levy2019electrodynamics,shanenko2008superconducting}
and has been commonly observed in several of our \ce{Al} electromigration
junctions. A more stable gate enables us to map the sample over a larger range
of $\epsilon_0$. For instance, the conductance map in the N state (Fig.
\ref{fig:sampleB}b) shows the zero-bias Kondo resonance developing at the
degeneracy point of the quantum dot (situated at $V_G^0 \approx 2$ V) and
reaching a minimum intensity at $V_G \approx -4$ V, indicating the center of the
odd diamond, and corresponding to the maximum of the YSR states' energy
dispersion in the $\pi$-phase. The complete concealment of the less intense
diamond edges, buried below the Kondo-enhanced ones, makes the analysis of the
diamond features and the extraction of $\alpha_G$ impossible in this sample.

According to the results obtained for the Kondo energy scale extraction in sample A, we give an experimental definition of $T_K$ as the FWHM/2.9 of the Kondo resonance,
this time subtracting the differential conductance background, taken as the
average conductance measured between V = $\pm 0.5$ to $\pm 1$ mV. By extracting
the YSR states' energies and replacing $V_G$ axis with the corresponding $T_K$
non-dimensionalized with the extracted $\Delta$, we find the occurrence of the
phase transition at $T_K/\Delta \approx 0.7$ (Fig. \ref{fig:sampleB}c),
analogous to sample A. This result points toward a similar ratio of $U/\Gamma$
in this sample, which is not striking since we employed the same nano-particles
colloidal solution in the fabrication of both samples, displaying a small size-distribution $>$10 \%
(leading to similar $U$ values). The nano-particles functionalizing ligands (dodecanethiol) may pose a limit for the minimum distance between the
nano-particle and the lead, or in other words, the tunnel coupling $\Gamma$. A
relatively higher Kondo-enhanced diamond edge w.r.t. the zero-bias resonance,
together with a sharp decrease of the YSR peak intensity across the phase
transition (Fig. \ref{fig:sampleB}d) may point towards a larger tunnel coupling
of the quantum dot to the probing electrode, becoming more sensitive to
differences in the refilling rate at both sides of the phase transition.

Faint resonances displaying the same energy dispersion and $V_G$ dependence of
the YSR peaks can be observed in Fig. \ref{fig:sampleB}a at higher energies,
appearing both inside the spectroscopic gap and also merging into the quasi-particle
continuum. A hypothetical spiky density of states above $\Delta_{probe}$ in the
probing lead (maybe caused by nanostructuration and granularity of the \ce{Al}
at the constriction) can result in low intensity replicas of the YSR resonances
at higher energies.




\end{document}